\documentclass{aa}
\usepackage{graphicx}
\begin{document}
\title{Mid- to Far-Infrared spectroscopy of Sharpless 171\footnote{Based on observations with ISO, an ESA project with instruments funded by ESA Member States (especially the PI countries: France, Germany, the Netherlands and the UK) and with the participation of ISAS and NASA.}}
\author{Y.  Okada \inst{1}
	\and
	T.  Onaka \inst{1}
	\and
	H.  Shibai \inst{2}
	\and
	Y.  Doi \inst{3}}
\institute{
Department of Astronomy, School of Science, University of Tokyo, Tokyo
113-0033, Japan
\and
Graduate School of Science, Nagoya University, Nagoya
464-8602, Japan
\and
Department of Earth Science and Astronomy, University of Tokyo, Tokyo
153-8902, Japan
}
\offprints{Y. Okada, \email{okada@astron.s.u-tokyo.ac.jp}}
\date{Received October 7, 2002; accepted September 11, 2003}

\abstract{We have collected one-dimensional raster-scan observations of the active star-forming region Sharpless 171 (S171), a typical \ion{H}{ii} region-molecular cloud complex, with the three spectrometers (LWS, SWS, and PHT-S) on board ISO.  We have detected 8 far-infrared fine-structure lines, [\ion{O}{iii}] 52\,$\mu$m, [\ion{N}{iii}] 57\,$\mu$m, [\ion{O}{i}] 63\,$\mu$m, [\ion{O}{iii}] 88\,$\mu$m, [\ion{N}{ii}] 122\,$\mu$m, [\ion{O}{i}] 146\,$\mu$m, [\ion{C}{ii}] 158\,$\mu$m, and [\ion{Si}{ii}] 35\,$\mu$m together with the far-infrared continuum and the H$_2$ pure rotation transition ($J=5$--3) line at 9.66\,$\mu$m.  The physical properties of each of the three phases detected, highly-ionized, lowly-ionized and neutral, are investigated through the far-infrared line and continuum emission.  Toward the molecular region, strong [\ion{O}{i}] 146\,$\mu$m emission was observed and the [\ion{O}{i}] 63\,$\mu$m to 146\,$\mu$m line ratio was found to be too small ($\sim 5$) compared to the values predicted by current photodissociation region (PDR) models.  We examine possible mechanisms to account for the small line ratio and conclude that the absorption of the [\ion{O}{i}] 63\,$\mu$m and the [\ion{C}{ii}] 158\,$\mu$m emission by overlapping PDRs along the line of sight can account for the observations and that the [\ion{O}{i}] 146\,$\mu$m emission is the best diagnostic line for PDRs.  We propose a method to estimate the effect of overlapping clouds using the far-infrared continuum intensity and derive the physical properties of the PDR.  The [\ion{Si}{ii}] 35\,$\mu$m emission is quite strong at almost all the observed positions.  The correlation with [\ion{N}{ii}] 122\,$\mu$m suggests that the [\ion{Si}{ii}] emission originates mostly from the ionized gas.  The [\ion{Si}{ii}] 35\,$\mu$m to [\ion{N}{ii}] 122\,$\mu$m ratio indicates that silicon of 30\% of the solar abundance must be in the diffuse ionized gas, suggesting that efficient dust destruction is undergoing in the ionized region.
}

\titlerunning{Infrared spectroscopy of Sharpless 171}
\maketitle

\section{Introduction}
Far-infrared (FIR) spectroscopy of the interstellar medium (ISM) provides us a great deal of information on the nature of the ISM.  For \ion{H}{ii} regions, FIR forbidden lines are useful tools to investigate the physical properties, such as the electron density and elemental abundance.  They are less subject to extinction and less sensitive to the electron temperature than optical forbidden lines.  Rubin et al. (\cite{Rubin}) described a semiempirical methodology to derive the electron density, the effective temperature for the ionizing star, and the gas-phase heavy element abundance from FIR line emissions based on the ionization bounded models.  In neutral regions, in which the gas is warmer than the molecular gas observed in radio frequencies, the kinetic temperature of atoms is high enough to excite FIR emission and a number of forbidden lines are emitted.  In the interface region between the ionized and molecular gas, intense far-ultraviolet (FUV)(6 eV $< h\nu <$ 13.6 eV) photons photodissociate molecules and photoionize heavy elements with ionization potential less than the Lyman limit.  This region is called the photodissociation region (PDR), where most energy is emitted in the FIR lines and continuum.  Theoretical models have been investigated for PDRs with various physical conditions (Tielens \& Hollenbach \cite{TH85}; Hollenbach et al. \cite{Hollenbach}; Wolfire et al. \cite{Wolfire}; Kaufman et al. \cite{Kaufman}; see Hollenbach \& Tielens \cite{HollenTielens} for a review).  These models take account of the chemistry and the energy balance and predict the cooling line intensities from PDRs.

In this paper, we report the results of a spectroscopic investigation of \object{Sharpless 171} (S171) with the Infrared Space Observatory (ISO; Kessler et al. \cite{Kessler}).  S171 is a large \ion{H}{ii} region and is the main cloud associated with the Cepheus OB4 stellar association (Yang \& Fukui \cite{YangFukui}).  The star cluster Be 59 located at the central portion of the nebulous region is the ionizing source of this region.  It contains one O7 star and several later-type stars.  The physical properties of the ionized gas have been investigated in radio continuum and recombination lines (e.g., Felli et al. \cite{Felli}; Rossano et al. \cite{Rossano}; Harten et al. \cite{Harten}).  Yang \& Fukui (\cite{YangFukui}) have mapped the large-scale molecular gas distribution with the \element[][13]{CO} ($J=1$--0) line, suggesting that Be 59 generates the ionization front on the surface of two dense molecular clumps (C1 and C2).  They suggested from the dynamics of the molecular clumps that the dense gas is contacting with the continuum source and the ionization front is driving shocks into the C1 clump.  S171 is a typical transition region from ionized gas to molecular clouds.  It has a scale of several tens arcminutes, which is appropriate for mapping observations in the FIR to study the physical properties of the ionized gas and PDR complex.

In Sect. 2, the observations and the data reduction are described.  The results are presented in Sect. 3 and discussed in Sect. 4.  A summary is given in Sect. 5.

\section{Observations and data reduction}

\begin{figure}[tb]
\centering
\resizebox{\hsize}{!}{\includegraphics{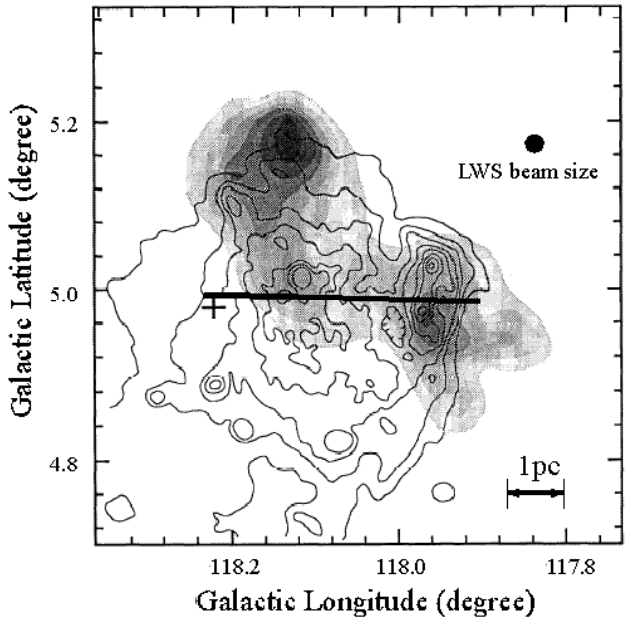}}
\caption{\element[][13]{CO} ($J=1$--0) integrated intensity map (in gray-scale) and the 0.61 GHz radio continuum emission map (in contours) (Yang \& Fukui \cite{YangFukui}).  The position of Be 59 is shown by the plus (+).  The thick solid line indicates the line along which the present observations were carried out.}
\label{target_position}
\end{figure}

%
\begin{figure}[t]
\centering
\resizebox{\hsize}{!}{\includegraphics{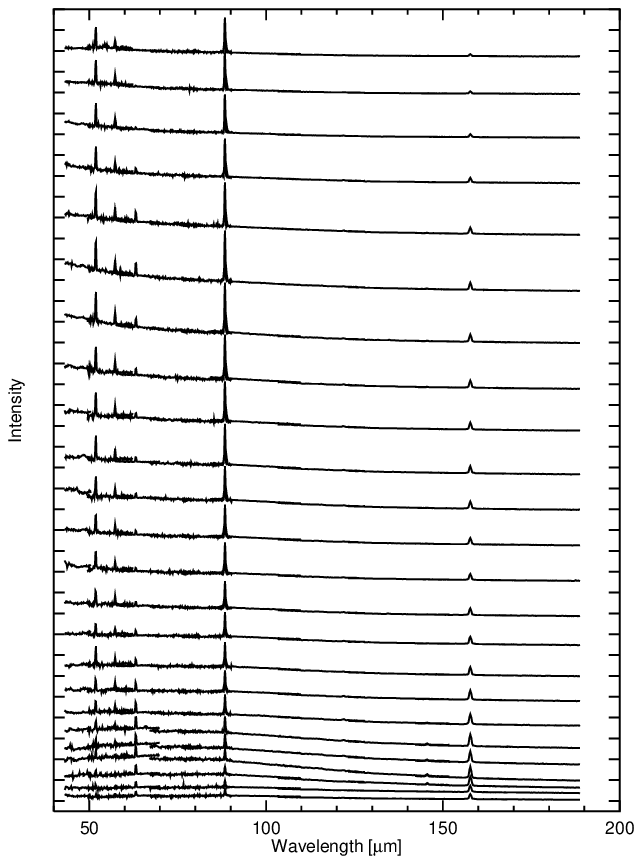}}
\caption{LWS spectra of the 24 raster positions.  The spectra are shifted in the $y$-axis for clarity from top (position 1) to bottom (position 24).  One division mark in the $y$-axis corresponds to $10^{-6}\,\mbox{W{\,}m}^{-2}\,\mu\mbox{m}^{-1}\mbox{sr}^{-1}$.}
\label{LWSspec}
\end{figure}

\begin{table*}
\rotatebox[origin=b]{90}{
\begin{minipage}{1.4\linewidth}
\caption{Summary of the line intensity of the LWS observations}
\label{res1}
\begin{center}
\begin{tabular}{ccccccccccc}
\hline
\multicolumn{4}{c}{position} &
\multicolumn{7}{c}{line intensities ($10^{-8}\,\mbox{W{\,}m}^{-2}\,\mbox{sr}^{-1}$)$^a$} \\ 
No. & $l$ (degree) & $b$ (degree) & $d$ (pc)$^b$ &  [\ion{O}{iii}] 52\,$\mu$m & [\ion{N}{iii}] 57\,$\mu$m$^c$ & [\ion{O}{i}] 63\,$\mu$m & [\ion{O}{iii}] 88\,$\mu$m & [\ion{N}{ii}] 122\,$\mu$m & [\ion{O}{i}] 146\,$\mu$m$^d$ & [\ion{C}{ii}] 158\,$\mu$m \\ \hline
 1 & 118.225 & 4.987 & 	0.14 & 31.50 $\pm$  2.88 & 10.73 $\pm$  1.26 &  $<$ 1.01 & 52.82 $\pm$  0.65 &  0.73 $\pm$  0.17 &  $<$ 0.24 &  6.49 $\pm$  0.26 \\  
 2 & 118.211 & 4.987 & 	0.15 & 35.51 $\pm$  3.67 & 13.17 $\pm$  1.23 &  $<$ 1.16 & 57.01 $\pm$  1.63 &  0.84 $\pm$  0.23 &  $<$ 0.20 &  7.37 $\pm$  0.26 \\  
 3 & 118.197 & 4.986 & 	0.33 & 32.90 $\pm$  4.84 & 14.25 $\pm$  1.13 &  2.84 $\pm$  0.55 & 59.16 $\pm$  1.13 &  1.23 $\pm$  0.19 &  0.31 $\pm$  0.16 &  9.44 $\pm$  0.36 \\ 
 4 & 118.183 & 4.986 & 	0.53 & 32.52 $\pm$  4.72 & 12.85 $\pm$  1.18 &  8.49 $\pm$  0.56 & 58.48 $\pm$  1.11 &  1.31 $\pm$  0.24 &  0.50 $\pm$  0.13 & 15.82 $\pm$  0.61 \\ 
 5 & 118.169 & 4.986 & 	0.74 & 43.68 $\pm$  2.93 & 15.99 $\pm$  1.17 & 15.93 $\pm$  0.63 & 65.29 $\pm$  1.15 &  1.39 $\pm$  0.36 &  0.97 $\pm$  0.18 & 20.86 $\pm$  0.88 \\ 
 6 & 118.155 & 4.986 & 	0.94 & 47.48 $\pm$  3.89 & 18.17 $\pm$  1.18 & 20.66 $\pm$  1.20 & 75.19 $\pm$  1.66 &  1.40 $\pm$  0.19 &  1.36 $\pm$  0.16 & 23.99 $\pm$  0.65 \\ 
 7 & 118.141 & 4.986 & 	1.15 & 43.30 $\pm$  5.03 & 17.64 $\pm$  0.79 & 13.64 $\pm$  1.25 & 75.05 $\pm$  1.96 &  1.73 $\pm$  0.32 &  1.10 $\pm$  0.13 & 24.13 $\pm$  0.82 \\ 
 8 & 118.127 & 4.985 & 	1.35 & 38.12 $\pm$  3.88 & 16.35 $\pm$  0.96 &  9.92 $\pm$  0.86 & 67.45 $\pm$  1.65 &  1.99 $\pm$  0.37 &  0.80 $\pm$  0.12 & 24.20 $\pm$  0.97 \\ 
 9 & 118.114 & 4.985 & 	1.56 & 38.10 $\pm$  3.41 & 14.97 $\pm$  1.02 &  7.83 $\pm$  0.61 & 61.18 $\pm$  1.72 &  2.20 $\pm$  0.24 &  0.65 $\pm$  0.12 & 23.72 $\pm$  0.87 \\ 
10 & 118.100 & 4.985 & 	1.76 & 34.52 $\pm$  3.88 & 13.82 $\pm$  1.04 &  7.82 $\pm$  0.38 & 59.02 $\pm$  1.87 &  1.84 $\pm$  0.23 &  0.49 $\pm$  0.13 & 23.80 $\pm$  0.99 \\ 
11 & 118.086 & 4.985 & 	1.97 & 31.66 $\pm$  3.44 & 12.05 $\pm$  1.00 &  7.16 $\pm$  0.58 & 49.63 $\pm$  1.92 &  1.54 $\pm$  0.23 &  0.56 $\pm$  0.10 & 24.28 $\pm$  1.08 \\ 
12 & 118.072 & 4.984 & 	2.17 & 30.08 $\pm$  1.52 & 12.36 $\pm$  0.69 &  6.54 $\pm$  0.41 & 46.47 $\pm$  2.75 &  1.55 $\pm$  0.24 &  $<$ 0.30 & 21.88 $\pm$  0.95 \\ 
13 & 118.058 & 4.984 & 	2.38 & 29.73 $\pm$  2.35 & 10.48 $\pm$  0.66 &  5.32 $\pm$  0.46 & 45.13 $\pm$  2.57 &  1.50 $\pm$  0.23 &  0.35 $\pm$  0.10 & 20.86 $\pm$  1.02 \\ 
14 & 118.044 & 4.984 & 	2.58 & 22.66 $\pm$  2.08 & 10.66 $\pm$  0.62 &  6.61 $\pm$  0.55 & 39.52 $\pm$  2.15 &  1.75 $\pm$  0.27 &  0.39 $\pm$  0.09 & 23.01 $\pm$  1.05 \\ 
15 & 118.030 & 4.984 & 	2.79 & 17.73 $\pm$  1.46 &  9.09 $\pm$  0.66 &  8.20 $\pm$  0.47 & 34.19 $\pm$  1.89 &  1.46 $\pm$  0.27 &  0.52 $\pm$  0.14 & 25.23 $\pm$  1.25 \\ 
16 & 118.016 & 4.984 & 	3.00 & 26.40 $\pm$  3.35 &  8.71 $\pm$  0.66 & 10.23 $\pm$  0.44 & 34.02 $\pm$  1.82 &  1.17 $\pm$  0.29 &  0.74 $\pm$  0.13 & 26.94 $\pm$  0.96 \\ 
17 & 118.002 & 4.983 & 	3.20 & 18.74 $\pm$  1.45 &  9.29 $\pm$  0.56 & 13.13 $\pm$  0.52 & 31.12 $\pm$  1.79 &  2.18 $\pm$  0.31 &  0.95 $\pm$  0.12 & 32.12 $\pm$  1.40 \\ 
18 & 117.988 & 4.983 & 	3.41 & 17.73 $\pm$  1.49 &  8.47 $\pm$  0.55 & 17.23 $\pm$  0.51 & 27.12 $\pm$  1.82 &  2.92 $\pm$  0.45 &  1.64 $\pm$  0.17 & 34.89 $\pm$  1.57 \\ 
19 & 117.974 & 4.983 & 	3.61 & 14.49 $\pm$  2.20 &  6.84 $\pm$  0.86 & 19.56 $\pm$  1.33 & 23.29 $\pm$  1.33 &  2.44 $\pm$  0.49 &  3.28 $\pm$  0.20 & 40.52 $\pm$  1.60 \\ 
20 & 117.960 & 4.983 & 	3.82 & 10.50 $\pm$  1.97 &  5.60 $\pm$  0.77 & 16.30 $\pm$  0.77 & 19.70 $\pm$  1.45 &  2.15 $\pm$  0.39 &  3.57 $\pm$  0.29 & 37.33 $\pm$  1.51 \\ 
21 & 117.946 & 4.982 & 	4.03 &  9.67 $\pm$  2.28 &  5.01 $\pm$  0.60 & 17.68 $\pm$  0.66 & 15.36 $\pm$  1.33 &  1.91 $\pm$  0.54 &  4.92 $\pm$  0.26 & 37.97 $\pm$  1.20 \\ 
22 & 117.932 & 4.982 & 	4.23 &  6.86 $\pm$  1.70 &  $<$ 1.70  & 14.46 $\pm$  0.93 & 12.01 $\pm$  0.86 &  1.01 $\pm$  0.32 &  3.44 $\pm$  0.33 & 32.48 $\pm$  1.97 \\ 
23 & 117.918 & 4.982 & 	4.44 &  6.19 $\pm$  1.46 &  2.50 $\pm$  0.51 &  8.58 $\pm$  0.67 & 10.09 $\pm$  0.85 &  0.73 $\pm$  0.16 &  0.82 $\pm$  0.11 & 22.00 $\pm$  1.11 \\ 
24 & 117.904 & 4.982 & 	4.64 &  3.15 $\pm$  1.30 &  3.25 $\pm$  0.57 &  6.35 $\pm$  0.57 &  9.23 $\pm$  1.04 &  0.85 $\pm$  0.16 &  0.79 $\pm$  0.25 & 17.21 $\pm$  0.75 \\ \hline
\end{tabular}
\end{center}
\begin{list}{}{}
\item[$^a$] Upper limit ($3\sigma$) is estimated from the statistical error of the baseline.
\item[$^b$] Distance from Be 59.
\item[$^c$] Weighted mean value of the intensities determined from the SW2 and SW3 channels.
\item[$^d$] Weighted mean value of the intensities determined from the LW3 and LW4 channels.
\end{list}
\end{minipage}}
\end{table*}

\begin{table*}
\centering
\caption{Summary of the line intensity of the SWS and PHT-S observations}
\begin{tabular}{ccccc}
\hline
& \multicolumn{4}{c}{line intensities ($10^{-8}\,\mbox{W{\,}m}^{-2}\,\mbox{sr}^{-1}$)$^a$} \\ 
No. & [\ion{Si}{ii}] 35\,$\mu$m & H$_2$ 9.66\,$\mu$m (SWS) & H$_2$ 9.66\,$\mu$m (PHT-S) & H$_2$  6.9\,$\mu$m \\ \hline
 1 &  $<$ 4.37  &  $<$ 3.02  & 	$<$  3.94 & $<$  2.52  \\
 2 &  $<$ 6.08  &  $<$ 2.20  & 	$<$  4.88 & $<$  5.62  \\
 3 &  $<$ 6.83  &  $<$ 2.22  & 	$<$  4.53 & $<$  3.80  \\
 4 &  $<$ 7.10  &  $<$ 2.44  & 	$<$  5.30 & $<$  3.82  \\
 5 &  $<$ 4.18  &  3.05 $\pm$  1.03 & 	$<$  6.16 & $<$  7.07  \\
 6 &  7.03 $\pm$  1.96 &  $<$ 2.74  & 	$<$  6.72 & $<$  6.08  \\
 7 &  6.20 $\pm$  1.68 &  $<$ 2.28  & 	$<$  6.80 & $<$  6.08  \\
 8 &  5.55 $\pm$  1.96 &  $<$ 2.19  & 	$<$  4.37 & $<$  5.81  \\
 9 &  7.24 $\pm$  1.40 &  $<$ 2.29  & 	$<$  3.51 & $<$  4.58  \\
10 &  5.73 $\pm$  1.81 &  $<$ 2.23  & 	$<$  3.61 & $<$  5.63  \\
11 &  7.25 $\pm$  1.99 &  $<$ 2.55  & 	$<$  4.71 & $<$  5.62  \\
12 &  $<$ 6.07  &  $<$ 2.26  & 	$<$  4.88 & $<$  4.89  \\
13 &  $<$ 6.75  &  $<$ 2.27  & 	$<$  3.85 & $<$  5.28  \\
14 &  $<$ 5.33  &  $<$ 2.00  & 	$<$  5.90 & $<$  7.04  \\
15 &  6.47 $\pm$  1.82 &  $<$ 3.64  & 	$<$  6.01 & $<$  5.98  \\
16 &  $<$ 5.76  &  $<$ 1.81  & 	$<$  4.45 & $<$  5.62  \\
17 &  7.97 $\pm$  1.72 &  $<$ 2.64  & 	$<$  5.30 & $<$  6.34  \\
18 & 10.62 $\pm$  1.71 &  $<$ 3.10  & 	$<$  4.19 & $<$  8.70  \\
19 & 17.76 $\pm$  1.92 &  2.81 $\pm$  0.87 & 	$<$  6.76 & $<$ 14.13  \\
20 & 11.05 $\pm$  1.65 &  5.82 $\pm$  0.63 & 	$<$  6.93 & $<$ 17.39  \\
21 & 18.62 $\pm$  2.20 &  8.34 $\pm$  1.09 & 	$<$  6.45 & $<$ 21.74  \\
22 &  7.86 $\pm$  1.79 &  3.30 $\pm$  0.97 & 	$<$  6.84 & $<$ 14.31  \\
23 &  $<$ 4.18  &  $<$ 3.87  & 	$<$  4.45 & $<$  7.97 \\
24 &  3.87 $\pm$  1.41 &  $<$ 2.37  & 	$<$  5.56 & $<$  7.61 \\\hline
\end{tabular}
\begin{list}{}{}
\item[$^a$] Upper limit ($3\sigma$) is estimated from the statistical error of the baseline for the SWS and from the uncertainty in the flux of the corresponding detector channel for the PHT-S
\end{list}
\label{res2}
\end{table*}

We made one-dimensional raster-scan observations of the S171 region with the Long-Wavelength Spectrometer (LWS; Clegg et al. \cite{Clegg}), the Short-Wavelength Spectrometer (SWS; de Graauw et al. \cite{deGraauw}), and the PHT Spectrometer (PHT-S; Lemke et al. \cite{Lemke}; Laureijs et al. \cite{Laureijs}) on board ISO.  The one-dimensional raster-scan started from the position near Be 59 to the region of $\sim 5$ pc from Be 59, which is spanned from the ionized gas to the molecular cloud (Fig.~\ref{target_position}).  The observations were carried out at 24 positions (Table~\ref{res1}), each of which was separated by $50\arcsec$.  The full grating scan mode AOT LWS01 was used in the LWS observations to cover the wavelength range from 43 to 197\,$\mu$m with $\lambda/\Delta\lambda\sim 100$--$300$ (Gry et al. \cite{Gry}).  Four grating scans were carried out for each raster position with the spectral sampling of a half of the resolution element.  The total integration time at a grating position was 2 sec.  The LWS beam size was $66\arcsec$--$86\arcsec$, depending on the wavelength, which corresponded to $0.27\sim 0.35$ pc at the distance of S171 of 850 pc (MacConnell \cite{MacConnell}), and the beam overlapped with each other.  The Off-Line Processing (OLP) version 10.1 data obtained from the ISO Archival Data Center were used for the present study.  The spectra were further processed by using the ISO Spectral Analysis Package (ISAP\footnote{The ISO Spectral Analysis Package (ISAP) is a joint development by the LWS and SWS Instrument Teams and Data Centers.  Contributing institutes are CESR, IAS, IPAC, MPE, RAL and SRON.}).  After removing remaining noises, the spectrum at each position was averaged over individual spectral scans.  The long wavelength part of the LWS spectra of extended sources is affected by the presence of fringes in the continuum (Swinyard et al. \cite{Swinyard}), therefore for $\lambda > 90$\,$\mu$m the spectra have been corrected for fringes using the procedure within ISAP.  Finally, the extended source correction was applied and the fluxes were converted into surface brightness according to the beam size of the detectors.  We used the latest values of the extended source correction factors and the beam sizes given in Gry et al. (\cite{Gry}).  The line intensities were obtained by Gaussian fits.  The absolute flux calibration uncertainty is reported to be 10--20\% for point sources and 50\% for extended sources (Gry et al. \cite{Gry}).  The relative accuracy is assumed to be 30\%, which was estimated from the gaps in the signal levels of the detectors.

The same positions were observed with the SWS and the PHT-S.  For the SWS observations the grating line profile mode AOT SWS02 was used to observe [\ion{Si}{ii}] 35\,$\mu$m and the H$_2$ pure rotational transition ($J=5$--3) at 9.66\,$\mu$m.  The aperture size was $20\arcsec\times 33\arcsec$ and $14\arcsec\times 20\arcsec$ and the spectral resolution was about 1500 and 2000 for the [\ion{Si}{ii}] 35\,$\mu$m and H$_2$ 9.66\,$\mu$m lines, respectively (Leech et al. \cite{Leech}).  The OLP 10.1 data were obtained from the ISO Archival Data Center and used in the present study except for the data at the position 23 in Table~\ref{res2}, for which only the processed data of version 9.1 were available.  At each raster position one up-down grating scan was carried out with the total integration time of 200 sec.  We used the ISAP for further data processing and obtained the line intensities by Gaussian fits.  The flux calibration accuracy is reported to be 7\% and 22\% for the H$_2$ 9.66\,$\mu$m and [\ion{Si}{ii}] 35\,$\mu$m lines, respectively, for point-like sources.  The conversion factors for extended sources are estimated to be accurate within 10\% (Leech et al. \cite{Leech}).

The PHT-S had a resolving power of $\lambda/\Delta\lambda \sim 90$ in two wavelength bands with the aperture size of $24\arcsec\times 24\arcsec$.  The data were reduced by the Phot-Interactive Analysis (PIA\footnote{The ISOPHOT data presented in this paper were reduced using PIA, which is a joint development by the ESA Astrophysics Division and the ISOPHOT Consortium with the collaboration of the Infrared Processing and Analysis Center (IPAC). Contributing ISOPHOT Consortium institutes are DIAS, RAL, AIP, MPIK, and MPIA.}) version 8.0.  The total integration time was 128 sec at each raster position.  The absolute flux calibration accuracy is about 10\% or less for point sources (Laureijs et al. \cite{Laureijs}).  Because of the low spectral resolution, we could not apply Gaussian fits to estimate the line intensity.  Instead, we simply measured the flux of the detector channel at the line center and subtracted the continuum estimated from the adjacent detectors.

\section{Results}

\begin{figure}
\centering
\resizebox{\hsize}{!}{\includegraphics{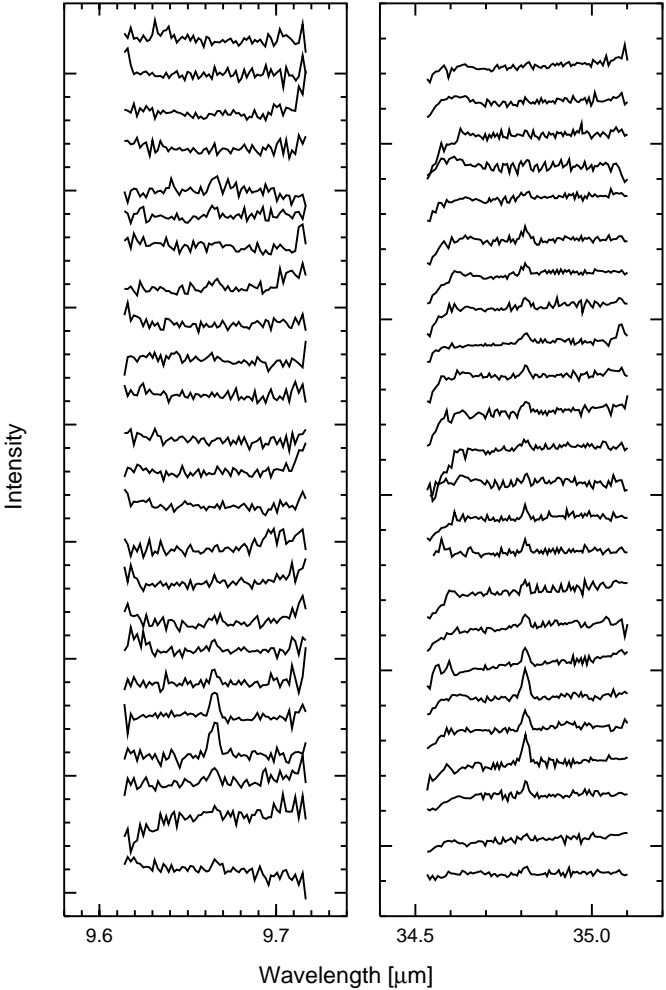}}
\caption{SWS spectra of the 24 raster positions.  The spectra are shifted in the $y$-axis for clarity from top (position 1) to bottom (position 24).  One division mark corresponds to $10^{-5}\,\mbox{W{\,}m}^{-2}\,\mu\mbox{m}^{-1}\mbox{sr}^{-1}$.}
\label{SWSspec}
\end{figure}

\begin{figure}
\centering
\resizebox{\hsize}{!}{\includegraphics{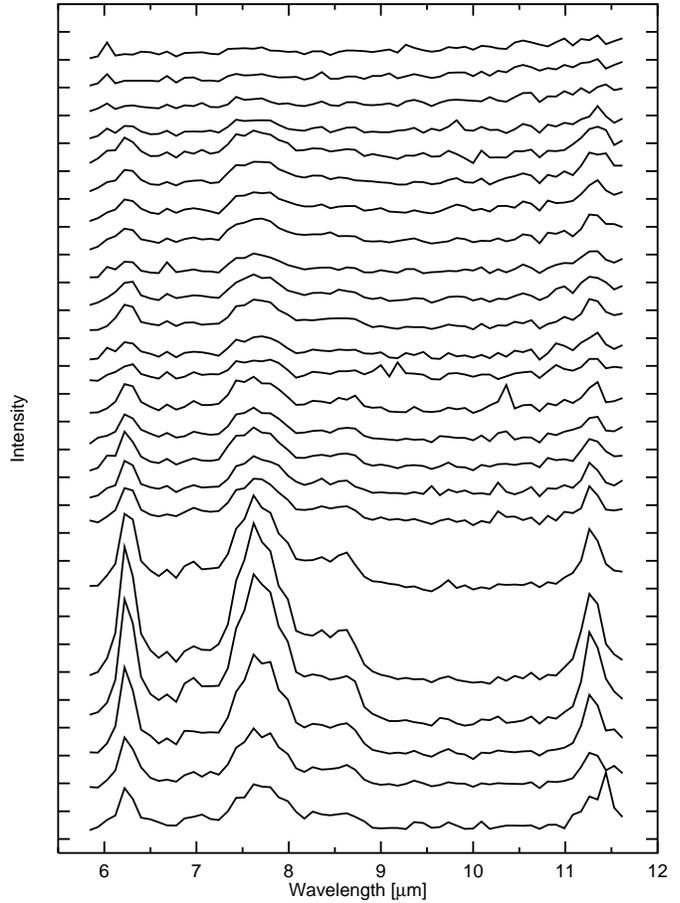}}
\caption{PHT-S spectra (5.6--11.6\,$\mu$m) of the 24 raster positions.  The spectra are shifted in the $y$-axis for clarity from top (position 1) to bottom (position 24).  One division mark corresponds to $10^{-6}\,\mbox{W{\,}m}^{-2}\,\mu\mbox{m}^{-1}\mbox{sr}^{-1}$.}
\label{PHTspec}
\end{figure}

Figures~\ref{LWSspec}, \ref{SWSspec}, and \ref{PHTspec} show the LWS, SWS, and PHT-S spectra of S171, respectively.  We detected 7 forbidden lines, [\ion{O}{iii}] 52\,$\mu$m, [\ion{N}{iii}] 57\,$\mu$m, [\ion{O}{i}] 63\,$\mu$m, [\ion{O}{iii}] 88\,$\mu$m, [\ion{N}{ii}] 122\,$\mu$m, [\ion{O}{i}] 146\,$\mu$m, and [\ion{C}{ii}] 158\,$\mu$m in the LWS spectra of almost all the observed positions.  The results of the LWS observations are summarized in Table~\ref{res1}.  The [\ion{N}{iii}] 57\,$\mu$m and [\ion{O}{i}] 146\,$\mu$m lines were observed by two adjacent detectors, SW2 and SW3, and LW3 and LW4, respectively.  The line intensities derived from different detectors are in agreement with each other within the estimated uncertainties.  We use the weighted mean of the intensities from the two detectors in the following analysis.  In the case that the line was detected only by one detector with the other being an upper limit, we adopted the intensity of the detected channel.  [\ion{O}{iii}] 88\,$\mu$m was also observed by two adjacent detectors, SW5 and LW1.  We used only the SW5 data for [\ion{O}{iii}] 88\,$\mu$m because of its higher spectral resolution.  We have detected [\ion{Si}{ii}] 35\,$\mu$m and the H$_2$ pure rotational transition at 9.66\,$\mu$m in the SWS observations, while H$_2$ 9.66\,$\mu$m and 6.9\,$\mu$m ($J=7$--5) were not detected at the 3$\sigma$ level in any PHT-S spectra.  Upper limits for these lines in the PHT-S spectra were estimated from the uncertainties in the flux of the corresponding detectors.  The upper limits of the 9.66\,$\mu$m line are compatible with the SWS results except for position 21, where the discrepancy may be attributed to the difference in the aperture size if the emission is highly peaked.  The results of the SWS and PHT-S observations are given in Table~\ref{res2}.  The errors of the line intensities include both the fitting errors and the statistical errors of the baseline for the LWS and SWS data.  Uncertainties in the absolute flux are not included in the errors in Tables~\ref{res1} and \ref{res2}.  For the line emission less than 3 times the uncertainty ($3\sigma$) we regard it as non-detection and give $3\sigma$ as the upper limit.

\begin{figure*}
\resizebox{\hsize}{!}{\includegraphics{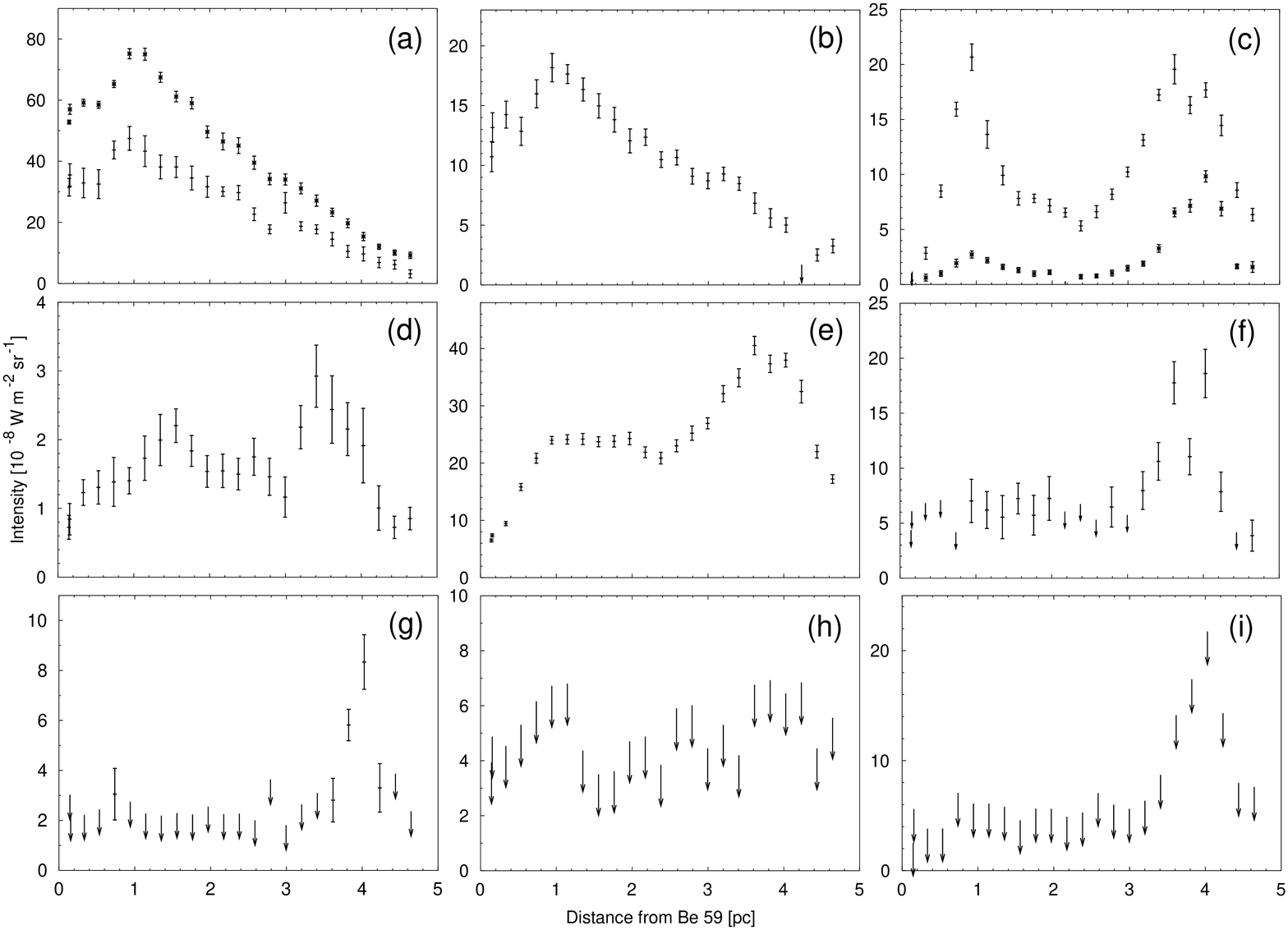}}
\caption{Observed line intensities against the distance from Be 59.  The arrows indicate the 3$\sigma$ upper limit.  {\bf a)} [\ion{O}{iii}] 52\,$\mu$m (lower points of +) and 88\,$\mu$m (upper points of $*$);  {\bf b)} [\ion{N}{iii}] 57\,$\mu$m; {\bf c)} [\ion{O}{i}] 63\,$\mu$m (upper points of +) and [\ion{O}{i}] 146\,$\mu$m multiplied by 2 (lower points of $*$);  {\bf d)} [\ion{N}{ii}] 122\,$\mu$m;  {\bf e)} [\ion{C}{ii}] 158\,$\mu$m;  {\bf f)} [\ion{Si}{ii}] 35\,$\mu$m;  {\bf g)} H$_2$ 9.66\,$\mu$m by the SWS;  {\bf h)} H$_2$ 9.66\,$\mu$m by the PHT-S;  and {\bf i)} H$_2$ 6.9\,$\mu$m.
}
\label{lines}
\end{figure*}

\begin{figure}
\centering
\resizebox{\hsize}{!}{\includegraphics{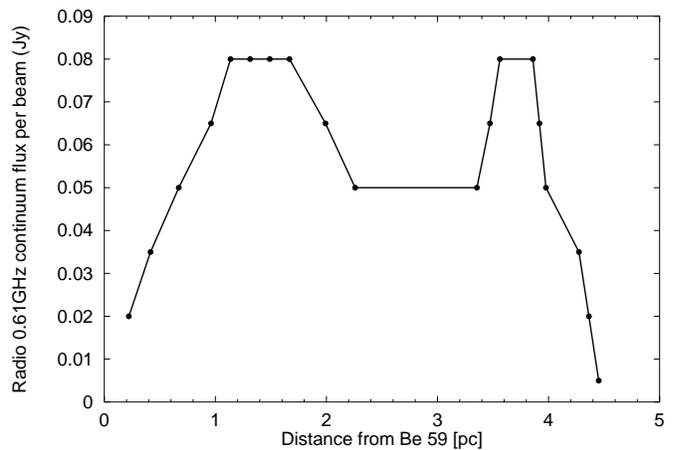}}
\caption{0.61GHz continuum flux (Harten et al. \cite{Harten}) on the line along which the present raster-scan observations were carried out.}
\label{0.61GHz}
\end{figure}

Figure~\ref{lines} shows the spatial distributions of the line intensities as a function of the distance from Be 59, $d$.  The [\ion{O}{i}] lines and the [\ion{N}{ii}] 122\,$\mu$m line show spatial distributions similar to the 0.61GHz radio continuum flux (Fig.~\ref{0.61GHz}).  The intensities of these lines show maxima at $d \sim 1$ pc and $\sim 4$ pc, indicating the presence of high density gas at these positions.  The [\ion{C}{ii}] 158\,$\mu$m, [\ion{Si}{ii}] 35\,$\mu$m, and H$_2$ 9.66\,$\mu$m emissions also show a strong peak at $\sim 4$ pc, but their intensity distributions become rather flat around $\sim 1$ pc.  The [\ion{O}{iii}] and [\ion{N}{iii}] lines appear strong at $d\sim 1$ pc and monotonically decrease with $d$, indicating a decrease in the abundance of the doubly ionized ions.

The coexistence of ionic lines of different ionization degrees as well as neutral atomic and molecular lines indicates that the observed region contains gases of various physical conditions.  In the following we attribute each observed line to three distinct physical phases, the highly-ionized gas, the lowly-ionized gas and the PDR gas, and investigate the properties of each gas.

\subsection{Highly-ionized gas}
\label{ssec:Hionized}

[\ion{O}{iii}] 52\,$\mu$m, 88\,$\mu$m and [\ion{N}{iii}] 57\,$\mu$m originate from the highly-ionized gas.  The [\ion{O}{iii}] 52\,$\mu$m to 88\,$\mu$m intensity ratio is in general a function of the electron density and temperature (Osterbrock \cite{Osterbrock}).  However, this ratio is insensitive to the electron temperature around $10^4$~K, and the electron density can be estimated directly from the observed ratio.  The maximum extinction in the observed positions is estimated as $\tau_{\mathrm{ext}}\sim 0.1$ at 52\,$\mu$m based on the \element[][12]{CO} line intensity at the C1 cloud (Yang \& Fukui \cite{YangFukui}) and the standard dust model (Draine \& Lee \cite{DraineLee}) with the conversion factor of $2.8\times 10^{20}$ H$_2$ molecule{\,}K$^{-1}\,\mbox{km}^{-1}\,\mbox{s}\,\mbox{cm}^{-2}$ (Mizutani et al. \cite{Mizutani}, for details).  Thus the extinction by dust does not affect the observed 52\,$\mu$m to 88\,$\mu$m ratio.

\begin{figure}
\resizebox{\hsize}{!}{\includegraphics{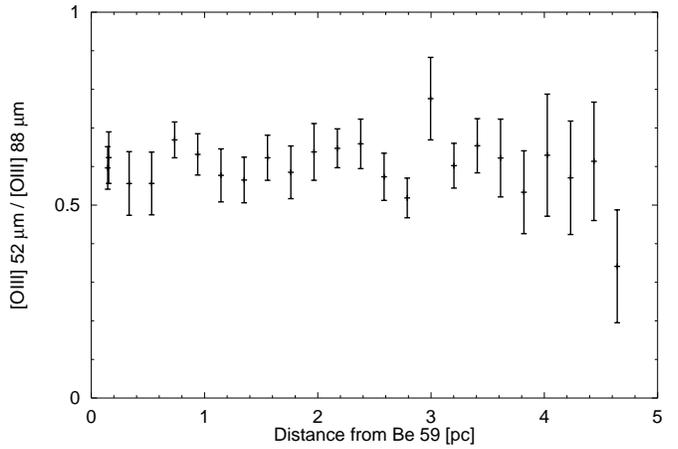}}
\caption{[\ion{O}{iii}] 52\,$\mu$m to 88\,$\mu$m ratio against the distance from Be 59.}
\label{obsOIIIratio}
\end{figure}

Figure~\ref{obsOIIIratio} shows that the observed ratio is approximately $\sim 0.6$ except for position 24 where the detection of [\ion{O}{iii}] 52\,$\mu$m is marginal.  The mean ratio 0.6 indicates $n_e \simeq 20$ cm$^{-3}$ for $T_e = 10^4$~K with the parameters for the transition given in Osterbrock (\cite{Osterbrock}).  The low density limit of the ratio is 0.56 for $T_e = 10^4$~K.  The uncertainty in the collision coefficients leads to an uncertainty of about 20\% in the ratio at the low density limit (Mizutani et al. \cite{Mizutani}).  The relative flux calibration error leads to an uncertainty of $\pm 0.2$ in the ratio.  Taking account of these uncertainties, the upper limit at the electron density is estimated as 150 cm$^{-3}$.  The lower limit is not constrained.  The derived range of the electron density ($n_e < 150$ cm$^{-3}$) is about the same as that in the Carina region ($n_e < 100$ cm$^{-3}$; Mizutani et al. \cite{Mizutani}), indicating the presence of the diffuse \ion{H}{ii} region also in S171.  According to Mizutani et al. (\cite{Mizutani}), the line-of-sight thickness of the line-emitting gas $L$ can be given by
\begin{equation}
L=1.2\times\left(\frac{I(\mathrm{[\ion{O}{iii}] 88\,}\mu\mathrm{m})}{10^{-7}\,\mathrm{W}\,\mathrm{m}^{-2}\,\mathrm{sr}^{-1}}\right)\times\left(\frac{20\,\mathrm{\,cm}^{-3}}{n_e}\right)^2\quad \mathrm{pc}\,,
\end{equation}
when $n_e$ is well below the critical density, the filling factor is unity, and $\log\mbox{(O/H)}=-3.3$ (Holweger \cite{Holweger}).  For the maximum [\ion{O}{iii}] 88\,$\mu$m intensity of $\sim 7.5\times 10^{-7}\,\mbox{W{\,}m}^{-2}\,\mbox{sr}^{-1}$, $L$ becomes approximately $9\times (20\mathrm{\,cm}^{-3}/n_e)^2$~pc.  This thickness is comparable with the projected size of S171 of several pc, suggesting that the highly-ionized gas is associated with S171.

The [\ion{N}{iii}] 57\,$\mu$m to [\ion{O}{iii}] 52\,$\mu$m intensity ratio is also insensitive to the temperature, and thus is a function of the electron density and the abundance ratio of N$^{++}$ and O$^{++}$ (Mizutani et al. \cite{Mizutani}).  N and O have similar second ionization potential energies and both lines can be assumed to come from the same region.  From the observed ratio of $\sim 0.4$ at most positions, we derive the abundance ratio N$^{++}$/O$^{++}$ as $0.23$ or $0.29$ for $T_e=10^4$~K and $n_e=20$~cm$^{-3}$ or $n_e=150$~cm$^{-3}$, respectively.  Taking account of the relative flux calibration error of the LWS, we estimate N$^{++}$/O$^{++}=0.2\pm 0.1$ for $n_e < 150$ cm$^{-3}$.  This is in agreement with the value of about 0.2 at the galactocentric distance $R_{\mathrm{gal}}=8.5$ kpc, which is estimated from 34 compact \ion{H}{ii} regions located between $R_{\mathrm{gal}}=0.3$ and 15 kpc (Mart\'{\i}n-Hern\'{a}ndez et al. \cite{MartinHernandez}), and is also similar to the value of $\sim 0.19$ of the core ionized gas in the Carina nebula (Mizutani et al. \cite{Mizutani}).

According to Rubin et al. (\cite{Rubin}) the effective temperature of the ionizing star can be estimated from the [\ion{N}{iii}] 57\,$\mu$m to [\ion{N}{ii}] 122\,$\mu$m line ratio for the ionization bounded condition.  With $T_e=10^4$~K and $n_e<150$~cm$^{-3}$, we derive $\langle\mathrm{N}^{++}\rangle$/$\langle\mathrm{N}^{+}\rangle\sim0.69$--$0.86$ from the total intensities of [\ion{N}{iii}] 57\,$\mu$m and [\ion{N}{ii}] 122\,$\mu$m of the 24 positions.  This indicates $T_{\mathrm{eff}}\sim37000$--$37500$~K of the ionization source, which is cooler of about 1000~K than that of the earliest type star in the S171 region, O7 ($T_{\mathrm{eff}}=38500$~K).  The [\ion{N}{ii}] 122\,$\mu$m emission comes also from the diffuse ionized gas and the ionization bounded condition may not hold.  Conversely an O7-type star can provide a sufficient amount of the [\ion{N}{iii}] 57\,$\mu$m emission to account for the observed intensity indicated by the [\ion{N}{ii}] 122\,$\mu$m emission.

\subsection{Neutral region}
\label{ssec:neutral}

The [\ion{O}{i}] 63\,$\mu$m, 146\,$\mu$m, [\ion{C}{ii}] 158\,$\mu$m, [\ion{Si}{ii}] 35\,$\mu$m, and H$_2$ lines originate from the PDR, where the FUV flux is converted into atomic and molecular gaseous line emissions by the photoelectric heating.  We estimate the physical properties from the observed intensities of the line and continuum emission, using the PDR model by Kaufman et al. (\cite{Kaufman}).  The major model parameters are the gas density, $n$, and the radiation field strength, $G_0$, in units of the solar neighborhood ($1.6\times 10^{-6}\,\mbox{W{\,}m}^{-2}$; Habing \cite{Habing}).  The model gives integrated line intensities seen in the face-on view.

\begin{figure}
\resizebox{\hsize}{!}{\includegraphics{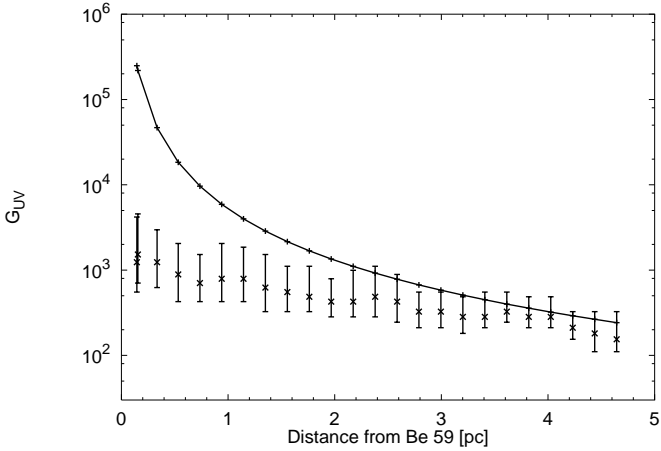}}
\caption{Radiation field intensity $G_{\mathrm{UV}}$ estimated from the luminosity of an O7 type star at the origin (solid line) and that from the temperature by fitting the LWS spectrum (crosses with errorbars, see text).}
\label{GUV}
\end{figure}

First, we fit the continuum emission of the LWS spectra with a gray-body radiation as
\begin{equation}
F(\lambda)=\tau_{100}\times \left(\frac{\lambda}{100\mu\mathrm{m}}\right)^{-\beta}\ B_{\lambda}(T_d)\,,
\end{equation}
where $T_d$ is the dust temperature, $\tau_{100}$ is the optical depth at 100\,$\mu$m, $B_\lambda$ is the Planck function, and $\beta$ is the power-law index of the emissivity on the wavelength.  We assume $\beta=1$ because it gives the best fit.  The obtained $T_d$ ranges from $34$~K to $53$~K and tends to decrease with the distance from Be 59.  The total far-infrared flux $FIR$ is given by
\begin{eqnarray}
FIR &=& \int_0^{\infty} F(\lambda)\,d\lambda\\
    &=& \tau_{100}\int_0^{\infty} \left(\frac{\lambda}{100\mu\mathrm{m}}\right)^{-\beta}\ B_{\lambda}(T_d)\ d\lambda\,.
\end{eqnarray}
In practice, the integration was carried out over the spectral range that makes effective contributions.  The far-infrared flux should be balanced by the absorbed energy in the ultraviolet to optical region as
\begin{eqnarray}
G_{\mathrm{UV}}\times (1.6\times 10^{-6} \mbox{\,W{\,}m}^{-2})&=& \int_{912\AA}^{\infty} 4\pi J^*_\lambda d\lambda\\
&=& \frac{4\pi}{\langle\tau_{\mathrm{abs}}\rangle} FIR \,\propto T^{\beta+4}\,,   \label{GUVeq}
\end{eqnarray}
where $J^*_\lambda$ is the mean intensity from the heating sources and $\langle\tau_{\mathrm{abs}}\rangle$ is the weighted mean of the absorption optical depth over $J_\lambda^*$.  We assume $\langle\tau_{\mathrm{abs}}\rangle/\tau_{100}=700$ (Okumura \cite{Okumura}; Chan et al. \cite{Chan}).  In Fig.~\ref{GUV} the crosses with errorbars show the derived $G_{\mathrm{UV}}$.  We also calculate $G_{\mathrm{UV}}$ as $L/4\pi d^2$, where $L$ ($=2.6\times 10^5 L_{\sun}$) is the luminosity of an O7 type star (solid line in Fig.~\ref{GUV}).  Both estimates are roughly in agreement with each other for $d>2$ pc.  In the region closer to the heating sources the estimate from the stellar luminosity becomes much larger.  It suggests that dust grains emitting the FIR emission are not present in the vicinity of Be 59.  The agreement in the outer region supports that Eq. (\ref{GUVeq}) gives a reasonable estimate for $G_{\mathrm{UV}}$.  In the PDR model the parameter $G_0$ does not include photons with the energy of $<6$~eV, which do not contribute to the photoelectric heating.  Thus $G_0$ is given by
\begin{equation}
G_0\times (1.6\times 10^{-6} \mbox{\,W{\,}m}^{-2})= \int_{912\AA}^{2066\AA} 4\pi J^*_\lambda d\lambda\,.
\end{equation}
For an O7 type star with $T_\mathrm{eff}=38000$~K, we estimate $G_{\mathrm{UV}}/G_0=1.3$ and use this scaling in the following analysis.

\begin{figure*}
\centering
\resizebox{0.8\hsize}{!}{\includegraphics{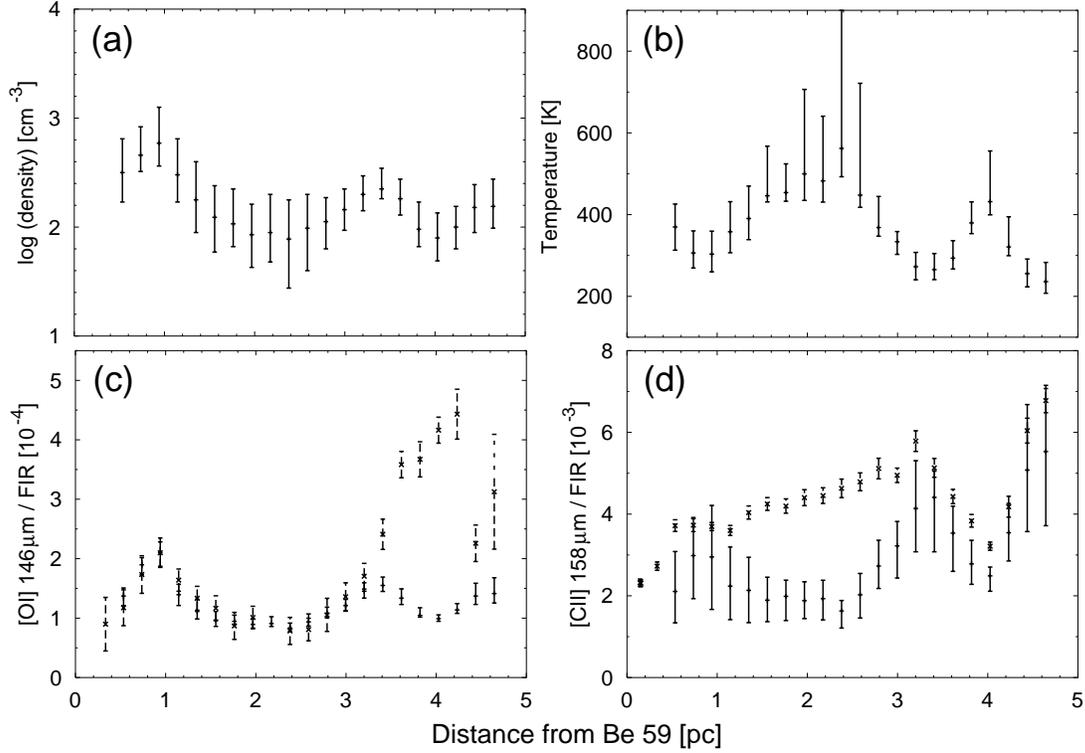}}
\caption{Results of the PDR model with [\ion{O}{i}] 63\,$\mu$m/$FIR$ as the input parameter (case 1).  {\bf a)} gas density, {\bf b)} surface temperature, {\bf c)} model prediction (solid errorbars) and observation (dashed errorbars) of [\ion{O}{i}] 146\,$\mu$m/$FIR$, and {\bf d)} model prediction (solid errorbars) and observation (dashed errorbars) of [\ion{C}{ii}] 158\,$\mu$m/$FIR$.  The errors of the model prediction include the uncertainties in the observed [\ion{O}{i}] 63\,$\mu$m intensities and $G_0$.}
\label{PDRi63}
\end{figure*}

\begin{figure*}
\centering
\resizebox{0.8\hsize}{!}{\includegraphics{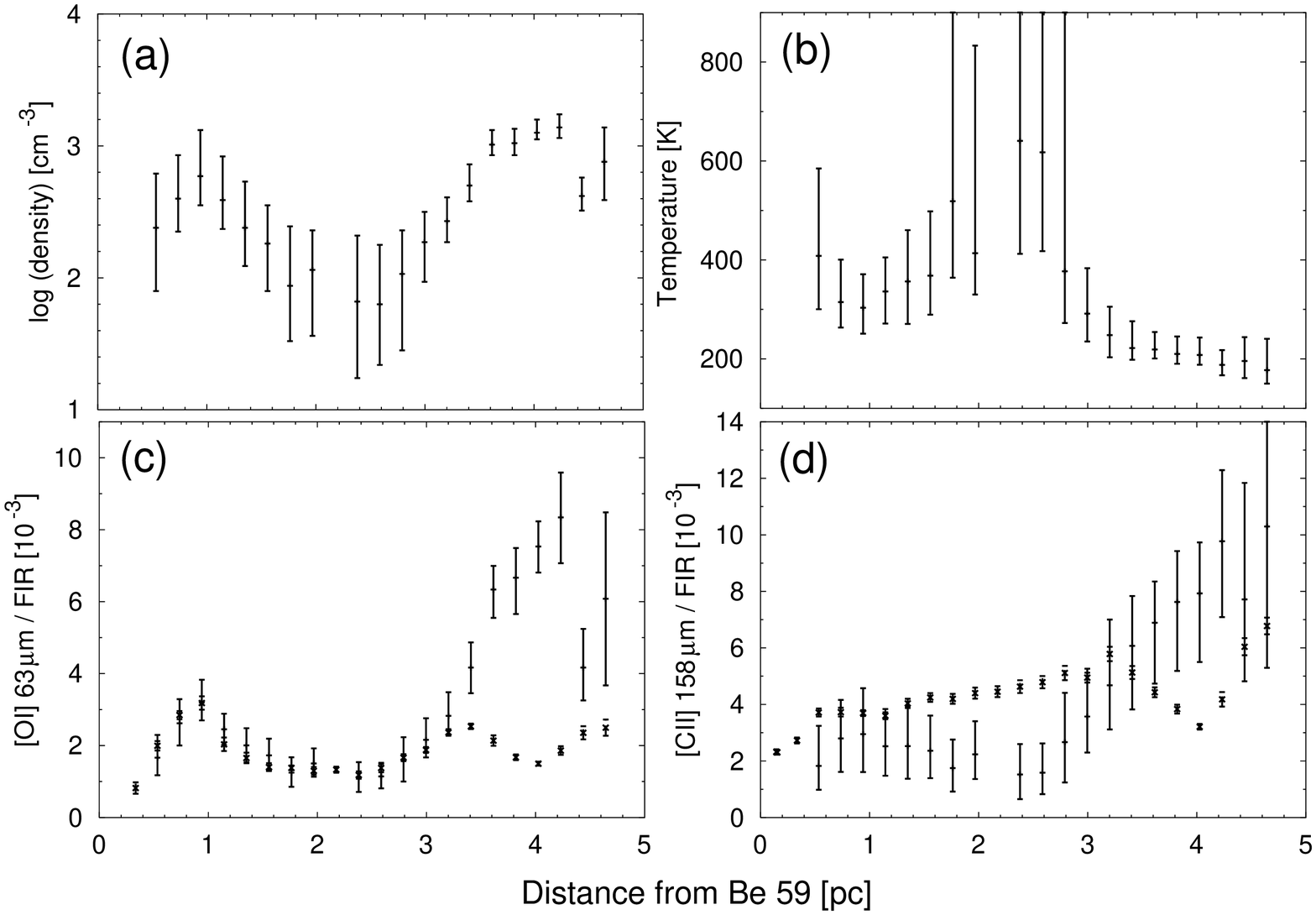}}
\caption{Results of the PDR model with [\ion{O}{i}] 146\,$\mu$m/$FIR$ as the input parameter (case 2).  {\bf a)} gas density, {\bf b)} surface temperature, {\bf c)} model prediction (solid errorbars) and observation (dashed errorbars) of [\ion{O}{i}] 63\,$\mu$m/$FIR$, and {\bf d)} model prediction (solid errorbars) and observation (dashed errorbars) of [\ion{C}{ii}] 158\,$\mu$m/$FIR$.  The errors of the model prediction include the uncertainties in the observed [\ion{O}{i}] 146\,$\mu$m intensities and $G_0$.}
\label{PDRi146}
\end{figure*}

[\ion{C}{ii}] 158\,$\mu$m and [\ion{O}{i}] 63\,$\mu$m and 146\,$\mu$m are the major cooling lines in the PDR.  However [\ion{C}{ii}] 158\,$\mu$m comes also from the ionized gas.  We estimate the density $n$ from $G_0$ and either of the ratio [\ion{O}{i}] 63\,$\mu$m/$FIR$ (case 1) or the ratio [\ion{O}{i}] 146\,$\mu$m/$FIR$ (case 2).  With the estimated $n$, we derive the ratio [\ion{C}{ii}] 158\,$\mu$m/$FIR$, the surface temperature, and the ratio [\ion{O}{i}] 146\,$\mu$m/$FIR$ or [\ion{O}{i}] 63\,$\mu$m/$FIR$, for cases 1 or 2, respectively.  The results are plotted in Figs.~\ref{PDRi63} and \ref{PDRi146}.  In both cases the model predicts fairly well for $d<3$~pc the intensity of the other [\ion{O}{i}] line that is not used for the input parameter, and always provides a smaller intensity of [\ion{C}{ii}] 158\,$\mu$m than observed.  The excess emission of [\ion{C}{ii}] 158\,$\mu$m is considered to come from the ionized gas.  At $d\sim 4$~pc, however, the observed [\ion{O}{i}] 146\,$\mu$m is much stronger than the model prediction in case 1 (Fig.~\ref{PDRi63}c), or the model predicts too strong [\ion{O}{i}] 63\,$\mu$m in case 2 (Fig.~\ref{PDRi146}c).  The model also predicts the [\ion{C}{ii}] 158\,$\mu$m intensity stronger than observed for $d>3.5$~pc (case 2; Fig.~\ref{PDRi146}d) or indicates that almost all the [\ion{C}{ii}] 158\,$\mu$m emission comes from the PDR gas for $d>3.5$~pc (case 1; Fig.~\ref{PDRi63}d).  The [\ion{N}{ii}] 122\,$\mu$m intensity shows a peak at $d\sim 4$~pc and thus there should be a significant contribution to [\ion{C}{ii}] 158\,$\mu$m from the ionized gas that emits [\ion{N}{ii}] 122\,$\mu$m.  Therefore the line intensities of [\ion{O}{i}] 63\,$\mu$m, 146\,$\mu$m and [\ion{C}{ii}] 158\,$\mu$m together with the FIR continuum emission can not be accounted for consistently by the latest PDR model in either case.  In case 2 the gas density has a peak at $\sim 4$~pc (Fig.~\ref{PDRi146}a), where CO observations indicate the presence of the main molecular clump C1 (Yang \& Fukui \cite{YangFukui}).  On the other hand case 1 indicates a local minimum in the density and a peak in the temperature at $\sim 4$~pc (Figs.~\ref{PDRi63}a and b), neither of which is compatible with the CO observations.  Case 2 leads to more consistent results with the CO observations (Figs.~\ref{PDRi146}a and b).  In the following we assume case 2 gives correct results and discuss the discrepancy in the [\ion{O}{i}] 63\,$\mu$m and [\ion{C}{ii}] 158\,$\mu$m emission.

According to the PDR model, [\ion{O}{i}] 63\,$\mu$m is optically thick, [\ion{C}{ii}] 158\,$\mu$m is marginally thick ($\tau\sim 1$), and [\ion{O}{i}] 146\,$\mu$m and the FIR emission are optically thin in a PDR (Kaufman et al. \cite{Kaufman}).  The largest value of the derived $\tau_{100}$ is $\sim 0.005$, supporting that the FIR emission is optically thin in the PDR of S171.  The optical depth effects of emission lines are taken into account in the PDR model.  If several PDR clouds overlap with each other along the line of sight, however, the optically thick lines will be attenuated by foreground PDRs, while optically thin emissions will be a simple summation of those from each PDR cloud.

The FIR intensity is estimated simply as $1.6\times 10^{-6} G_{\mathrm{UV}}/4\pi\ (\mathrm{W\,m}^{-2}\mathrm{\,sr}^{-1})$ in the PDR model by assuming that all the incident radiation is absorbed and converted into the FIR emission (Kaufman et al. \cite{Kaufman}).  Here we define the overlapping factor $Z$ as
\begin{equation}
Z=\frac{4\pi FIR}{1.6\times 10^{-6}\,(\mathrm{W\,m}^{-2}\mathrm{\,sr}^{-1})\, G_{\mathrm{UV}}}\,.    \label{Zdef}
\end{equation}
For $Z<1$, $Z$ is supposed to indicate the beam filling factor, whereas it should estimate the degree of overlapping of clouds for $Z>1$ since the FIR emission is optically thin.

\begin{figure}
\resizebox{\hsize}{!}{\includegraphics{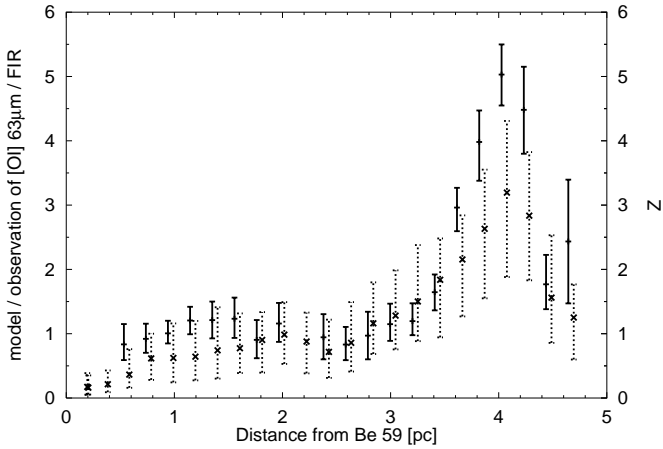}}
\caption{Ratio of the model prediction (case 2) to the observations for the [\ion{O}{i}] 63\,$\mu$m/FIR ratio (solid errorbars) and the overlapping factor $Z$ (dotted errorbars; Eq. (\ref{Zdef})).  The overlapping factors are shifted slightly to the right for clarity.}
\label{compOI}
\end{figure}

\begin{figure}
\resizebox{\hsize}{!}{\includegraphics{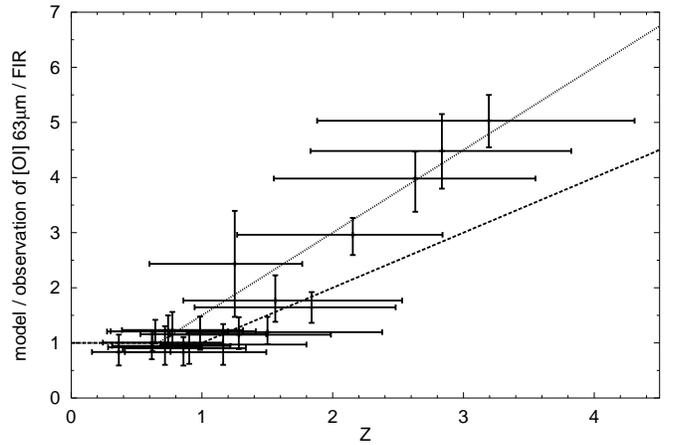}}
\caption{Ratio of the model prediction (case 2) to the observation of [\ion{O}{i}] 63\,$\mu$m/FIR against the overlapping factor $Z$.  The dashed line indicates the relation calculated by Eq. (\ref{compsum_eq}) with $\tau\gg 1$, while the dotted line indicates that shifted by $-50$\% in $Z$.}
\label{compOIsum}
\end{figure}

Figure~\ref{compOI} plots the spatial variation of the ratio of the model prediction (case 2) to the observation for the [\ion{O}{i}] 63\,$\mu$m/$FIR$ ratio and $Z$.  The two quantities are plotted in Fig.~\ref{compOIsum}, indicating a good correlation.  $FIR$ and [\ion{O}{i}] 146\,$\mu$m are not affected by overlapping because both emissions are optically thin ($\tau\ll 1$), and we assumed that the results for case 2 can directly be used.  The PDR model predicts the observed [\ion{O}{i}] 63\,$\mu$m/$FIR$ ratio fairly well ($\mbox{model/observation}\sim 1$) for the positions with $Z<1$.  Figure~\ref{compOI} shows that the regions of $d>3.5$~pc have $Z$ larger than unity, where the discrepancy between the model and the observations is significant (see also Figs.~\ref{PDRi146}c and d).  The value of $Z$ larger than unity indicates more than one clouds in the beam, suggesting that absorption of the overlapping clouds is significant for the [\ion{O}{i}] 63\,$\mu$m emission.

Assuming that $N$ PDR clouds, each of which has the same optical depth $\tau$, overlap on the line of sight, the total observed intensity normalized by the intensity from a cloud can be given by
\begin{equation}
\sum_{j=0}^{N-1}e^{-j\tau}=\frac{1-e^{-N\tau}}{1-e^{-\tau}}\,.   \label{compsum_eq}
\end{equation}
We assume $\tau\gg 1$ for [\ion{O}{i}] 63\,$\mu$m.  The line of Eq. (\ref{compsum_eq}) with $\tau\gg 1$ is also plotted in Fig.~\ref{compOIsum} by substituting $N$ by the overlapping factor $Z$.  Figure~\ref{compOIsum} indicates that the correlation can be accounted for by Eq. (\ref{compsum_eq}) if $Z$ is shifted by $-50$\% (dotted line).  Such a shift can be attributed to the uncertainty in the absolute scale of $FIR$ and/or to the simple assumption of the model.

\begin{figure}
\resizebox{\hsize}{!}{\includegraphics{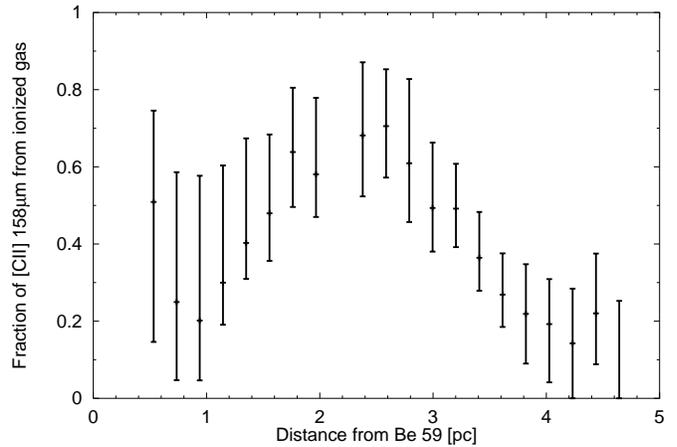}}
\caption{Estimated fraction of [\ion{C}{ii}] 158\,$\mu$m from the ionized gas.  The errorbars include the uncertainties in the observed intensities of [\ion{C}{ii}] 158\,$\mu$m and [\ion{O}{i}] 146\,$\mu$m, and $G_0$.}
\label{CIIfrac}
\end{figure}

For [\ion{C}{ii}] 158\,$\mu$m, we assume $\tau=1$ and apply Eq. (\ref{compsum_eq}) with the shift of $-50$\% in $Z$ to the model results predicted from $G_0$ and [\ion{O}{i}] 146\,$\mu$m/$FIR$.  The discrepancy that the model intensity is stronger than the observations for [\ion{C}{ii}] 158\,$\mu$m for $d>3.5$~pc, is resolved by taking account of the absorption due to the overlapping clouds.  The resultant model prediction for the [\ion{C}{ii}] 158\,$\mu$m emission becomes less than those observed at all the positions.  Then the excess intensity in the observed [\ion{C}{ii}] 158\,$\mu$m over the overlapping model prediction can be ascribed to that from the ionized gas.  Figure~\ref{CIIfrac} plots the thus estimated fraction of the [\ion{C}{ii}] 158\,$\mu$m emission from the ionized gas, indicating that a non-negligible fraction (20--70\%) of the [\ion{C}{ii}] 158\,$\mu$m emission comes from the ionized gas.  The amount of the ionized gas contribution to [\ion{C}{ii}] 158\,$\mu$m is within a reasonable range expected from the [\ion{N}{ii}] 122\,$\mu$m intensity (see Sect. \ref{ssec:Lionized}).  Thus the simple model of overlapping PDR clouds accounts for the observations consistently.

\subsection{Lowly-ionized gas}
\label{ssec:Lionized}

\begin{figure}
\resizebox{\hsize}{!}{\includegraphics{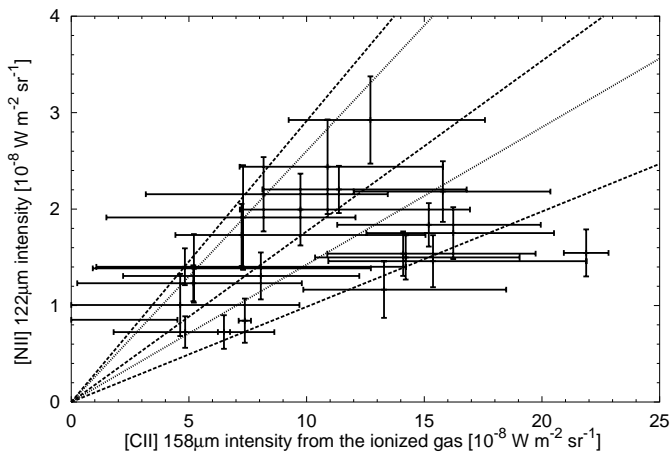}}
\caption{Observed [\ion{N}{ii}] 122\,$\mu$m intensity versus estimated [\ion{C}{ii}] 158\,$\mu$m intensity from the ionized gas.  The dashed lines indicate the calculations for the solar abundance gas of $n_e=70$, $30$, and $10$ cm$^{-3}$ (from the upper to lower lines).  The dotted lines show those for the ISM abundance gas of $n_e=10$ cm$^{-3}$ (upper line) and at the low density limit (lower line).  The electron temperature is assumed to be $10^4$~K for all cases.}
\label{NIICII}
\end{figure}

[\ion{N}{ii}] 122\,$\mu$m, and part of [\ion{C}{ii}] 158\,$\mu$m and [\ion{Si}{ii}] 35\,$\mu$m originate from the lowly-ionized gas.  Using the observed [\ion{N}{ii}] 122\,$\mu$m intensity and the [\ion{C}{ii}] 158\,$\mu$m intensity estimated from the ionized gas in Sect. \ref{ssec:neutral}, we derive the electron density by assuming the abundance ratio of C$^+$ to N$^+$.  Model calculations suggest that in the warm low-density photoionized interstellar medium, 81\% of nitrogen is in N$^+$ and 95\% of carbon is in C$^+$ (Sembach et al. \cite{Sembach}).  The difference of the ionization fractions is smaller than that in the estimate of the [\ion{C}{ii}] 158\,$\mu$m intensity from the ionized gas and we simply assume that the ratio of C$^+$ to N$^+$ is equal to the gas phase abundance ratio of C to N.  Figure~\ref{NIICII} plots the observed [\ion{N}{ii}] 122\,$\mu$m intensity against the [\ion{C}{ii}] 158\,$\mu$m intensity from the ionized gas.  We assume two cases for the abundance, the latest solar abundance ($\log\mbox{(N/H)}=-4.07$ and $\log\mbox{(C/H)}=-3.41$; Holweger \cite{Holweger}) and the ISM abundance ($\log\mbox{(N/H)}=-4.10$ and $\log\mbox{(C/H)}=-3.86$; Savage \& Sembach \cite{SavageSembach}).  Figure~\ref{NIICII} suggests that the observed [\ion{N}{ii}] 122\,$\mu$m and [\ion{C}{ii}] 158\,$\mu$m intensities can arise from the ionized gas with the electron density of about $30^{+40}_{-20}$~cm$^{-3}$ for the solar abundance and that of less than 10\,cm$^{-3}$ for the ISM abundance for $T_e=10^4$~K.  COBE observations suggest that the electron density of the [\ion{N}{ii}] 122\,$\mu$m emitting gas in the Galactic plane is typically 10--30\,cm$^{-3}$ (Wright et al. \cite{Wright}), which is in agreement with the present estimate.  We estimate the path length of the N$^+$ gas in the same manner as for the O$^{++}$ gas.  With the strongest [\ion{N}{ii}] 122\,$\mu$m intensity of $2.9\times 10^{-8}$\,W\,m$^{-2}$\,sr$^{-1}$ and assuming $T_e=10^4$~K and the solar abundance, we derive the length $L=4.2$ and $0.9$ pc for $n_e=30$ and $70$ cm$^{-3}$, respectively.  Because of the large uncertainty in $n_e$ due to the uncertainty in the abundance, it is difficult to conclude whether the lowly-ionized gas is distributed in a thin sheet or extending over the S171 region.

\begin{figure}
\resizebox{\hsize}{!}{\includegraphics{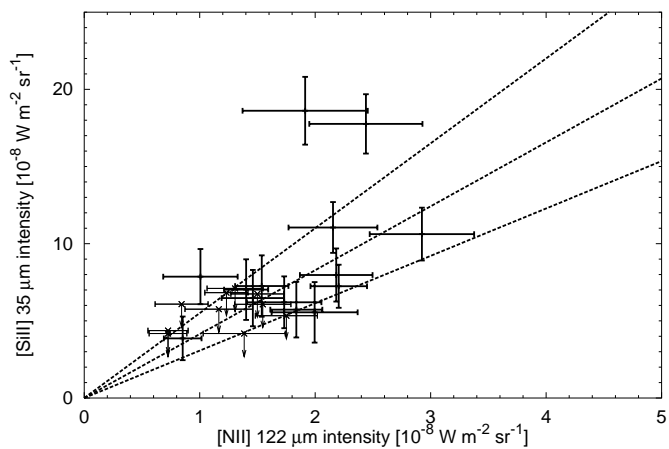}}
\caption{Observed [\ion{N}{ii}] 122\,$\mu$m intensity versus [\ion{Si}{ii}] 35\,$\mu$m intensity.  The upper line indicates the gas of the Si abundance of 40\% of solar and the low density limit.  The middle line indicates the gas of the Si abundance of 30\% of solar and $n_e=30$ cm$^{-3}$.  The lower line indicates the gas of the Si abundance of 20\% of solar and $n_e=70$ cm$^{-3}$.  The electron temperature is assumed to be $10^4$~K for all the cases.}
\label{NIISiII}
\end{figure}

The observed [\ion{N}{ii}] 122\,$\mu$m and [\ion{Si}{ii}] 35\,$\mu$m intensities give us an estimate of the gas phase silicon abundance.  The N abundance is assumed to be solar without depletion and the singly ionized ion fraction is assumed to be the same for N and Si.  Both lines are also assumed to originate from the same ionized region.  Silicon and nitrogen have quite different ionization potential energies.  However, model calculations indicate that 81\% of nitrogen is in N$^+$ and 91\% of silicon is in Si$^+$ in the warm ionized medium (WIM) owing to the charge exchange effect (Sembach et al. \cite{Sembach}).  Taking account of the calculated ionization fractions the required silicon abundance is lowered by about 10\% compared to the assumption of the same ionization fraction.  Figure~\ref{NIISiII} shows the observed intensities and the results of the calculation.  S171 shows strong silicon emission, which requires $\log\mbox{(Si/H)}=-4.99$, $30\pm 10$\% of solar abundance silicon in the gas phase for $n_e<70$ cm$^{-3}$.

Figure~\ref{NIISiII} shows that the [\ion{Si}{ii}] 35\,$\mu$m emission is well correlated with the [\ion{N}{ii}] 122\,$\mu$m emission except for the two positions where the [\ion{Si}{ii}] 35\,$\mu$m emission is the strongest.  This correlation suggests that the [\ion{Si}{ii}] 35\,$\mu$m emission originates mainly from the [\ion{N}{ii}] 122\,$\mu$m emitting gas.  Observations of the Carina region by Mizutani et al. (\cite{Mizutani_pro}) also suggest a good correlation between [\ion{Si}{ii}] 35\,$\mu$m and [\ion{N}{ii}] 122\,$\mu$m, indicating that the [\ion{Si}{ii}] 35\,$\mu$m emission comes mostly from the ionized gas.  The strong [\ion{Si}{ii}] 35\,$\mu$m emission of the two points, $d\sim 4$ pc, may be attributed to the contribution from the PDR or the high electron density gas.

The [\ion{Si}{ii}] 35\,$\mu$m line has a high critical density ($2.6\times 10^5$ cm$^{-3}$) for the H$^0$ collision and a high exciting temperature ($413$~K).  Hence, in the neutral region the [\ion{Si}{ii}] 35\,$\mu$m emission indicates the presence of high-temperature and high-density gas.  In Sect. \ref{ssec:neutral}, the density of the PDR in S171 is estimated to be 100--1500 cm$^{-3}$ and the temperature is $200$~K--$400$~K.  These properties indicate that the [\ion{Si}{ii}] 35\,$\mu$m emission from the PDR should not be significant as Fig.~\ref{NIISiII} indicates.  Based on the comparison of the observed [\ion{Si}{ii}] 35\,$\mu$m emission with the PDR model by Hollenbach et al. (\cite{Hollenbach}), we found that we need by about 12 times larger silicon abundance in the gas than that assumed in the PDR model (about 2\% of solar abundance) if the strongest [\ion{Si}{ii}] 35\,$\mu$m emission ($d\sim 4$~pc) comes only from the PDR.  The [\ion{Si}{ii}] 35\,$\mu$m emission should also be attenuated by overlapping clouds because of its large optical depth.  Taking account of this effect, we estimate that silicon of $65\pm 20$\% of solar abundance must be in the gas phase in the PDR at $\sim 4$~pc.  If a half of the [\ion{Si}{ii}] 35\,$\mu$m emission at $\sim 4$~pc comes from the ionized gas and the other half from the PDR, then both regions need gaseous silicon of about 30\% of solar abundance.  Therefore the large Si abundance in the gas in the S171 region is a secure conclusion.

Intense [\ion{Si}{ii}] emission has been reported in several \ion{H}{ii} regions (Haas et al. \cite{Haas}), in the Galactic center region (Stolovy et al. \cite{Stolovy}), and in the Carina region (Mizutani et al. \cite{Mizutani_pro}).  Silicon and carbon are major constituents of interstellar grains (Mathis \cite{Mathis}).  The gas phase Si abundance is about 5\% of solar in cool clouds (Savage \& Sembach \cite{SavageSembach}) and the depleted atoms are thought to reside in interstellar grains.  The suggested large abundance of gaseous silicon in PDRs or \ion{H}{ii} regions indicates efficient grain destruction processes taking place in these regions.  Observations of the massive star-forming region the Carina nebula indicate the ratio of [\ion{Si}{ii}] 35\,$\mu$m to [\ion{N}{ii}] 122\,$\mu$m of about 8 (Mizutani et al. \cite{Mizutani_pro}), while it is about 4 in S171. In both regions, the estimated electron density is in a similar range (10--50\,cm$^{-3}$) and thus the difference in the ratio can be attributed to the difference in the relative abundance.  Observations of the starburst galaxy NGC 253 also indicate that about a half of the solar abundance Si should be in the gas (Carral et al. \cite{Carral}).  The present result indicates a slightly smaller value of $\sim 30$\%.  The degree of destruction may indicate the difference in the activity of the regions.  Theoretical investigations suggest that dust grains are destroyed quite efficiently by supernova shocks in a time scale of $2$--$4 \times 10^8$ yr (Jones et al. \cite{Jones94}, \cite{Jones96}).  Observations by the Goddard High Resolution Spectrograph (GHRS) on board the Hubble Space Telescope have indicated that Si returns to the gas phase relatively quickly, while Fe grains seem to be more resistive (Fitzpatrick \cite{Fitzpatrick96}, \cite{Fitzpatrick97}).  The present observations support these results and indicate that Si in dust grains is easily returned to the gas phase.   This is in contrast to the constant gas abundance suggested for oxygen and carbon (Cardelli et al. \cite{Cardelli}; Sofia et al. \cite{Sofia}; Cartledge et al. \cite{Cartledge}).  The recycling as well as the composition of dust grains in interstellar space should be reexamined based on these recent results of the gas abundance in various phases (cf. Tielens \cite{Tielens}; Jones \cite{Jones00}; Onaka \cite{Onaka}).

\section{Discussion}
\subsection{The absorption of [\ion{O}{i}] 63\,$\mu$m and [\ion{C}{ii}] 158\,$\mu$m}
\label{ssec:disc_OI}

In the S171 region, strong [\ion{O}{i}] 146\,$\mu$m emission is observed compared to [\ion{O}{i}] 63\,$\mu$m at $\sim 4$~pc, where the ratio of [\ion{O}{i}] 63\,$\mu$m/146\,$\mu$m is about $5$.  Small ratios compared to the PDR model prediction have been reported also in other objects.  Liseau et al. (\cite{Liseau}) showed that the ratio becomes 1--5 in the $\rho$ Oph cloud and were not able to identify any convincing mechanisms to explain the discrepancy.  They took account of the collisional excitation at low temperatures and examined the collision coefficients of O$^0$ in detail.  The model prediction based on the newly-derived collision parameters, however, did not make a significant difference from the older one.  Thus the uncertainty in the collision coefficients is not a major factor for the low [\ion{O}{i}] 63\,$\mu$m to 146\,$\mu$m ratio.  The $^3$P$_1$ and $^3$P$_0$ levels in \ion{O}{i} can be populated by cascades from higher states that have been populated from $^3$P$_2$ through the absorption of the UV interstellar radiation (Keenan et al. \cite{Keenan}).  With the oscillator strengths given by Morton (\cite{Morton}) we calculated the detailed balance of \ion{O}{i} level populations including this effect. The UV pumping is effective at low hydrogen densities (e.g. $n_{\mathrm{H}} < 1000$~cm$^{-3}$ and $T<1000$~K) and the [\ion{O}{i}] 63\,$\mu$m to 146\,$\mu$m ratio becomes smaller than 15 for temperatures of several hundred K.  Taking account of the possible optical depth effect, this process can account partly for the obtained low ratio.  However, the low ratio is observed at the surface of the molecular region, while the UV pumping is expected to be effective in the vicinity of the star.  The spatial distribution of the line ratio suggests that the UV pumping is not a major mechanism for the low ratio.  As described in Sect. \ref{ssec:Hionized}, dust extinction does not affect the observed line intensity for $\lambda > 50\,\mu$m and thus does not affect the observed [\ion{O}{i}] line ratio.  Caux et al. (\cite{Caux}) claimed that the low value of [\ion{O}{i}] 63\,$\mu$m to 146\,$\mu$m in the $\rho$ Oph cloud is due to the presence of a very large column density of atomic oxygen, which makes the 63\,$\mu$m line optically thick.  Giannini et al. (\cite{Giannini}) showed that the complex site of massive star formation, NGC 2024 (Orion B, W12), has the [\ion{O}{i}] 63\,$\mu$m to 146\,$\mu$m ratio of about 5 and suggested that the [\ion{O}{i}] 63\,$\mu$m line is strongly absorbed by the cold foreground gas.  Absorption of the [\ion{O}{i}] 63\,$\mu$m line has been reported directly by high spectral resolution KAO observations toward star-forming regions (Poglitsch et al. \cite{Poglitsch}; Kraemer et al. \cite{Kraemer}), and by ISO/LWS observations toward Sagittarius B2 (Baluteau et al. \cite{Baluteau}) and W49N (Vastel et al. \cite{Vastel}). 

Carbon atoms are in CO in cold molecular clouds, and thus the [\ion{C}{ii}] 158\,$\mu$m emission will not be absorbed in cool clouds as efficiently as [\ion{O}{i}] 63\,$\mu$m.  However, [\ion{C}{ii}] 158\,$\mu$m is marginally optically thick, $\tau\sim 1$, in the PDR (Kaufman et al. \cite{Kaufman}).  Boreiko \& Betz (\cite{Boreiko95}) observed NGC 6334 at high spectral resolution with the KAO and indicated a self-absorption profile of the [\ion{C}{ii}] 158\,$\mu$m emission at several positions.  NGC 6334 is a star-forming region where self-absorption of [\ion{O}{i}] 63\,$\mu$m is also observed (Kraemer et al. \cite{Kraemer}).  Boreiko \& Betz (\cite{Boreiko97}) further directly detected the absorption feature of [\ion{C}{ii}] 158\,$\mu$m in the star-forming region, NGC 3576, and suggested that the [\ion{C}{ii}] 158\,$\mu$m emission is absorbed by a foreground cloud.  The present analysis indicates that the model of overlapping PDR clouds accounts for observations sufficiently well.  The resultant [\ion{C}{ii}] 158\,$\mu$m emission from the ionized gas is in the reasonable range, supporting the present simple model.  A detailed study of the $\rho$ Oph region, where the contribution to [\ion{C}{ii}] 158\,$\mu$m from the ionized gas is negligible and thus the observed line intensities can be more directly compared to model predictions, also suggests that the [\ion{O}{i}] 63\,$\mu$m, 146\,$\mu$m, and [\ion{C}{ii}] 158\,$\mu$m intensities are able to be accounted for consistently by the present model with the overlapping factor (Okada et al. 2003).

These results suggest that the absorption of [\ion{O}{i}] 63\,$\mu$m and [\ion{C}{ii}] 158\,$\mu$m in clouds on the line of sight is the most likely cause for the difference between the PDR model predictions and the present observations.  The overlapping model does not necessarily assume separate clumpy clouds, but can also be applied if a thick cloud is viewed in an edge-on configuration.  In the PDR model by Kaufman et al. (\cite{Kaufman}), the turbulent velocity dispersion is assumed to be 1.5\,km\,s$^{-1}$, which corresponds to the line width of 2.5\,km\,s$^{-1}$ in FWHM.  In S171, the $^{13}$CO emission seems to have two components, one with the width of 1.0\,km\,s$^{-1}$ and the other with 3.4\,km\,s$^{-1}$ width (Yang \& Fukui \cite{YangFukui}).  Thus if the clouds are overlapping, self-absorption in [\ion{O}{i}] 63\,$\mu$m and [\ion{C}{ii}] 158\,$\mu$m can be expected to some extent in S171.  The validity of the present model of overlapping PDRs can be examined in detail by high spectral-resolution observations.  The present analysis suggests that the best diagnostic line for PDRs is [\ion{O}{i}] 146\,$\mu$m, and that $FIR$ is an important parameter to estimate the degree of the overlapping on the line of sight.

\subsection{Comparison with observations of other \ion{H}{ii} regions}

Observations of compact \ion{H}{ii} regions suggest that their electron density is in the range 200--10000 cm$^{-3}$ (Moorwood et al. \cite{Moorwood}; Watson et al. \cite{Watson}; Mart\'{\i}n-Hern\'{a}ndez et al. \cite{MartinHernandez}).  Observations of G29.96-0.02 suggest the presence of a dense ($n_\mathrm{e} \sim 57000$ cm$^{-3}$) core and a diffuse component ($n_\mathrm{e} \sim 680$ cm$^{-3}$; Morisset et al. \cite{Morisset}).  The \ion{H}{ii} region-molecular cloud complex of NGC 2024 has been shown to have $n_\mathrm{e}=$ 1200--1500 cm$^{-3}$ (Giannini et al. \cite{Giannini}).  The present results of S171 indicate the lower end of the electron density in the range of these observations, suggesting the presence of the diffuse \ion{H}{ii} region in S171.  This is supported by the extended distribution of the [\ion{O}{iii}] and [\ion{N}{iii}] lines over several pc.  Observations of the Carina nebula also indicate the presence of the low-density ($< 100$ cm$^{-3}$) highly-ionized gas extending over several tens pc (Mizutani et al. \cite{Mizutani}).  The presence of low-density highly-ionized gas in star-forming regions may not be an uncommon phenomenon.

In the Carina nebula, however, the [\ion{O}{i}] 63\,$\mu$m emission is weak and is about one third of the [\ion{C}{ii}] 158\,$\mu$m emission.  The [\ion{O}{i}] 146\,$\mu$m emission is also weak and has been detected only at several positions (Mizutani et al. \cite{Mizutani_pro}).  Compared to the Carina nebula, S171 has a large neutral gas density.  On the other hand, observations of NGC 2024 suggest a much higher density ($\sim 5 \times 10^5$--$10^6$ cm$^{-3}$) for the PDR (Giannini et al. \cite{Giannini}).  S171 may be at the middle stage in between dense and young star-forming regions and evolved diffuse PDRs.

The fraction of the [\ion{C}{ii}] emission from the ionized gas is suggested to be less than 30\%  both in the NGC 2024 and the Carina regions.  The present observations suggest a higher ratio of 20--70\% of the [\ion{C}{ii}] emission that comes from the ionized gas.  This ratio has a large uncertainty because it depends on the overlapping model, but if real, the large fraction may be partly due to the geometrical effect: in S171 the ionized gas and PDR overlap on the line of sight with a large degree.  Observations of the gas motion should provide useful information to further investigate the geometry of the S171 region.

\section{Summary}

We observed S171 with the LWS, the SWS, and the PHT-S on board ISO.  We have detected 7 forbidden lines, [\ion{O}{iii}] 52\,$\mu$m, [\ion{N}{iii}] 57\,$\mu$m, [\ion{O}{i}] 63\,$\mu$m, [\ion{O}{iii}] 88\,$\mu$m, [\ion{N}{ii}] 122\,$\mu$m, [\ion{O}{i}] 146\,$\mu$m, and [\ion{C}{ii}] 158\,$\mu$m, and far-infrared continuum with the LWS, [\ion{Si}{ii}] 35\,$\mu$m and H$_2$ 9.66\,$\mu$m with the SWS.  The PHT-S observations gave only upper limits on the H$_2$ 9.66\,$\mu$m and 6.9\,$\mu$m lines.  We attribute the observed lines to three phases -- highly-ionized gas, lowly-ionized gas, and PDR gas -- and derived the physical properties of each phase.

For the highly-ionized gas, we have estimated the electron density from the [\ion{O}{iii}] lines as $n_e < 150$ cm$^{-3}$.  The path length of the O$^{++}$ gas along the line of sight is estimated to be $9\times (20\mathrm{\,cm}^{-3}/n_e)^2$ pc, indicating that the highly-ionized gas is associated with S171.  The abundance ratio of N$^{++}$/O$^{++}$ is $0.2\pm 0.1$, in agreement with the observed ratio of \ion{H}{ii} regions at the galactocentric distance of about 8~kpc.

The [\ion{O}{i}] 63\,$\mu$m, 146\,$\mu$m and [\ion{C}{ii}] 158\,$\mu$m line intensities around the interface region between the ionized and molecular gases cannot be accounted for consistently by the current PDR model.   The [\ion{O}{i}] 146\,$\mu$m emission is very strong, and the [\ion{O}{i}] 63\,$\mu$m to 146\,$\mu$m ratio is too small.  The [\ion{C}{ii}] 158\,$\mu$m emission also seems to be too weak.  We conclude that the model based on the observed [\ion{O}{i}] 146\,$\mu$m/$FIR$ ratio gives the most consistent results with the other observations.  We propose a model in which the [\ion{O}{i}] 63\,$\mu$m and [\ion{C}{ii}] 158\,$\mu$m emissions are attenuated by overlapping PDR clouds.  We introduce the overlapping factor $Z$ defined as the ratio of the observed $FIR$ to the far-infrared intensity of a PDR cloud estimated from the temperature of the FIR continuum emission.  The observed intensities of the three lines can be accounted for by the simple model of overlapping PDRs.  We obtain the gas density $n\sim 100$--$1500$~cm$^{-3}$ and the temperature $T\sim 200$--$400$~K.  The present analysis indicates that the [\ion{O}{i}] 63\,$\mu$m line emission must be used with caution in diagnostics of PDRs and the [\ion{O}{i}] 146\,$\mu$m line may be the best parameter for the diagnostics. A comparison of the observations with the model prediction indicates that 20--70\% of the [\ion{C}{ii}] 158\,$\mu$m comes from the ionized gas.

We have obtained the electron density $n_e=30^{+40}_{-20}$ cm$^{-3}$ for the solar abundance or less than 10 cm$^{-3}$ for the ISM abundance for the lowly-ionized gas.  [\ion{Si}{ii}] 35\,$\mu$m is quite strong and the observed ratio of [\ion{Si}{ii}] 35\,$\mu$m to [\ion{N}{ii}] 122\,$\mu$m indicates the silicon abundance in the diffuse ionized gas to be about 30\% of solar.  This value is much larger than the typical value in interstellar diffuse clouds ($\sim$ 5\% of solar), suggesting the efficient dust destruction in the ionized region.  The present observations indicate the importance of the [\ion{Si}{ii}] line in future observations to investigate the dust destruction and the circulation processes of the interstellar medium.

\begin{acknowledgement}
The authors thank all the members of the Japanese ISO group, particularly H. Okuda,  K. Kawara, and Y. Satoh for their help during the observations and continuous encouragement.  We also thank the referee, Dr. L. Spinoglio, for many useful suggestions that greatly improved the paper.  This work was supported in part by Grant-in-Aids for Scientific Research from the Japan Society for the Promotion of Science (JSPS).  The authors thank K. Mochizuki for useful discussions.
\end{acknowledgement}

\end{document}